\documentclass[conference, a4paper]{IEEEtran}
\usepackage{cite}
\usepackage{subcaption} 
\usepackage{amsmath,amssymb,amsfonts}
\usepackage{graphicx}
\usepackage{booktabs} 
\usepackage{siunitx}
\usepackage{tabularray}
\usepackage{textcomp}
\usepackage{tikz}
\usepackage{amsmath}
\usepackage{amsfonts}
\usepackage{adjustbox}
\usepackage{multirow}
\usepackage{xcolor}
\usepackage{pgfplots}
\usepackage{url}
\usepackage{multirow} 

\usepackage{subcaption}
\usepackage[affil-it]{authblk} 
\usepackage{booktabs}
\usepackage{algpseudocode}
\usepackage[ruled, lined, longend, linesnumbered]{algorithm2e}
\usepackage[utf8]{inputenc}
\usepackage{tikz}

\usepackage{amssymb}
\usepackage{pifont}
\begin{document}

\title{AI-Driven Fast and Early Detection of IoT Botnet Threats: A Comprehensive Network Traffic Analysis Approach}

\author[1,2]{Abdelaziz Amara korba}
\author[1]{Aleddine Diaf}
\author[2]{Yacine Ghamri-Doudane}

\affil[1]{LRS, Badji Mokhtar University of Annaba, Algeria}
\affil[2]{L3I, University of La Rochelle, France}

\maketitle
\begin{abstract}
In the rapidly evolving landscape of cyber threats targeting the Internet of Things (IoT) ecosystem, and in light of the surge in botnet-driven Distributed Denial of Service (DDoS) and brute force attacks, this study focuses on the early detection of IoT bots. It specifically addresses the detection of stealth bot communication that precedes and orchestrates attacks. This study proposes a comprehensive methodology for analyzing IoT network traffic, including considerations for both unidirectional and bidirectional flow, as well as packet formats. It explores a wide spectrum of network features critical for representing network traffic and characterizing benign IoT traffic patterns effectively. Moreover, it delves into the modeling of traffic using various semi-supervised learning techniques. Through extensive experimentation with the IoT-23 dataset—a comprehensive collection featuring diverse botnet types and traffic scenarios—we have demonstrated the feasibility of detecting botnet traffic corresponding to different operations and types of bots, specifically focusing on stealth command and control (C2) communications.The results obtained have demonstrated the feasibility of identifying C2 communication with a 100\% success rate through packet-based methods and 94\% via flow-based approaches, with a false positive rate of 1.53\%.

\end{abstract}


\section{Introduction}

In the digital era, the proliferation of Internet of Things (IoT) devices has spurred unprecedented advancements across various sectors, from enhancing home automation to revolutionizing industrial processes. However, this rapid expansion has also exposed new vulnerabilities, positioning these interconnected networks as prime targets for sophisticated cyber threats. Among these, botnet attacks have emerged as a particularly significant challenge \cite{Nokia2023ThreatReport}. A botnet, a network of infected devices controlled by a malicious actor, can silently compromise numerous devices to orchestrate disruptions like Distributed Denial of Service (DDoS) attacks, as seen with the Mirai attack against OVH. The 2023 Nokia Threat Intelligence Report \cite{Nokia2023ThreatReport} and FortiGuard Labs' 2022 \cite{lee2016internet} review collectively underline the escalating cyber threat landscape, marked by a surge in IoT botnet-driven DDoS traffic and brute force attacks. This includes a significant increase in IoT device involvement, from 200,000 to 1 million within a year, and a rise in mobile trojans compromising banking data.

Botnet operations unfold in several stages: scanning, infection, control, and finally, attack. Mirai \cite{chris}, a notorious bot malware, exemplifies the botnet lifecycle, which encompasses four phases: scanning, infection, control, and attack. During the scanning phase, Mirai searches for vulnerable IoT devices with open Telnet ports. Once identified, the infection phase begins, exploiting weak default credentials to compromise devices. In the control phase, infected devices, now bots, connect to a Command and Control (C\&C) server, receiving instructions. Finally, in the attack phase, these bots execute coordinated attacks, such as DDoS. Much research has focused on detecting botnet-led attacks, particularly the final phase, which often involves volumetric DDoS attacks. These attacks are markedly distinct from legitimate traffic due to their network flow volume. To effectively prevent such attacks and ensure early detection, our study delves into identifying network traffic  associated with the stages preceding the attack phase. Detecting network traffic related to the scanning and infection stages allows for interrupting the process and preventing device compromise. Moreover, identifying command and control (C2) communication aids in detecting IoT devices compromised by botnets. This proactive approach aims to neutralize the bot malware before it reaches the attack stage, and mitigating their propagation. 

The landscape of IoT botnet detection has been extensively researched, with numerous studies focusing on the development of network-based intrusion detection systems (NIDS) that utilize artificial intelligence (AI). Despite the impressive performance of existing research in detecting botnet attacks, a significant gap remains in their prevention and early-stage detection. Most current methodologies are centered on recognizing and mitigating botnet attacks after their occurrence, overlooking the crucial need for proactive and anticipatory measures to prevent these threats before they materialize. This situation underscores a broader issue within the field: a notable deficiency in the early detection of IoT bots.

Furthermore, reducing the detection delay is critically important, both in the event of actual attacks and in the stages leading up to them. Minimizing this delay is vital to limit the impact of the infection and prevent the spread of bot malware throughout the network. Nonetheless, only a few existing studies have concentrated on reducing this response time. Our study aims to minimize the detection delay by conducting a thorough investigation of the representation of network traffic (flows/packets) and the features that characterize it.

Most state-of-the-art solutions predominantly use supervised learning methods for detecting botnet attacks. This approach, however, presupposes the availability of botnet traffic for training, an assumption not consistently valid in real-world scenarios, which often exhibit a significant imbalance between benign and malicious traffic. Furthermore, this dependence substantially restricts the model's capability to identify unknown botnet traffic, thus reducing its effectiveness against new or evolving threats. Our study intends to explore semi-supervised learning methods that do not necessitate malicious traffic for training.

Building on the identified gaps in existing research, several challenges need to be addressed to enhance botnet early detection and prevention capabilities. Firstly, due to the scarcity of malicious traffic for training purposes, there is a critical need for detection approaches that either do not require or require minimal malicious traffic for model training. Accurate recognition of normal traffic patterns is essential for detecting botnet activities. However, differentiating normal traffic from malicious traffic, which corresponds to the stealthy communication of bot malware---specifically during the infection and command and control (C2) phases---can be challenging, especially if the latter closely mimics benign behavior. Furthermore, it becomes even more challenging to not only detect but also reduce the detection delay effectively.

To address the challenge of detecting botnet activities with minimal or no access to malicious traffic, this study explores the potential of semi-supervised learning techniques, focusing on one-class classification methods, to accurately model normal network behavior. This innovative approach aims to uncover a wide spectrum of botnet traffic, including previously unknown bots, by identifying deviations from established network traffic patterns. Our research assesses the feasibility of 5 semi-supervised techniques in modeling benign network patterns and detecting a wide range of bot types. To detect stealth communications, such as infections and C2 traffic, this study examines several representations of network traffic, including bidirectional flows, unidirectional flows, and packets, along with the network features that characterize them. We explore and utilize features based on packet headers, without analyzing the payload of network packets. This method facilitates the detection of bot malware that uses encryption to hide its traffic, a feature increasingly critical in an era where encryption is commonly employed to bypass detection. By conducting a comprehensive evaluation of traffic representation, modeling methodologies, and sampling strategies, our study achieves a significant reduction in detection delays, aiming for less than one second. Our experimental validation with the IoT-23 dataset, featuring authentic traffic from IoT devices and a diverse array of verified IoT bot malware, underscores the efficacy of our methodology. We achieve a a high detection rate exceeding for scan traffic and C2 communications from various bot malwares, with a detection delay of 1 second in flow-based and less than one second in packet-based detection. 

The remainder of this paper is organized as follows. Section~\ref{RT} describes related work. The proposed methodology is presented in Section~\ref{SOL}. Section~\ref{SIM} depicts the performance evaluation results, and finally, Section~\ref{CON} concludes the paper.

\section{Related Work} \label{RT}
The increasing sophistication of IoT botnets necessitates advanced detection methodologies that not only identify threats but do so promptly to mitigate potential damage. Recent research in this domain has paved the way for various innovative detection strategies, each contributing uniquely to the field's advancement. This section reviews these contributions, particularly emphasizing the evolution towards early-stage and rapid detection mechanisms.

Wei et al. \cite{wei2023lightweight} developed a deep learning framework targeting early-stage IoT botnet detection, notable for its use of a 5-second detection window. This approach leverages packet payload-independent features, marking a significant step towards accurate and timely identification of network anomalies associated with botnet activities. Nguyen et al. \cite{nguyen2022collaborative} explored the potential of collaborative machine learning models in the early detection of IoT botnets, assessing various algorithms such as Support Vector Machine and K-Nearest Neighbors. Their work contributes valuable insights into the effectiveness of machine learning techniques in identifying botnet threats at an incipient stage. In a related effort, Nguyen et al. \cite{nguyen2022advanced} enhanced detection methodologies through a hybrid model that integrates PSI-rooted subgraph features, focusing on combining static and dynamic analysis to improve detection precision. While their model advances the detection capabilities, it primarily emphasizes the identification rather than the swift response to IoT botnet threats. Bojarajulu et al. \cite{bojarajulu2023intelligent} proposed a novel optimization strategy, SMIE (Slime Mould with Immunity Evolution), to optimize a hybrid classifier comprising Bidirectional Gated Recurrent Units (Bi-GRU) and Recurrent Neural Networks (RNN). This approach signifies an important development in enhancing the accuracy of botnet detection mechanisms.

Despite these advancements, a gap remains in the field for an approach that integrates the benefits of early detection with the requisite speed to respond to threats effectively. The present study aims to fill this gap by proposing a detection methodology that not only identifies IoT botnets at an early stage but does so with a significantly reduced detection time. Our approach is designed to offer a rapid response capability, crucial for limiting the impact of botnet attacks on IoT systems, thereby advancing beyond the current state-of-the-art in both detection timeliness and efficiency.

\begin{figure*}[]
    \centering
    \includegraphics[scale=0.45]{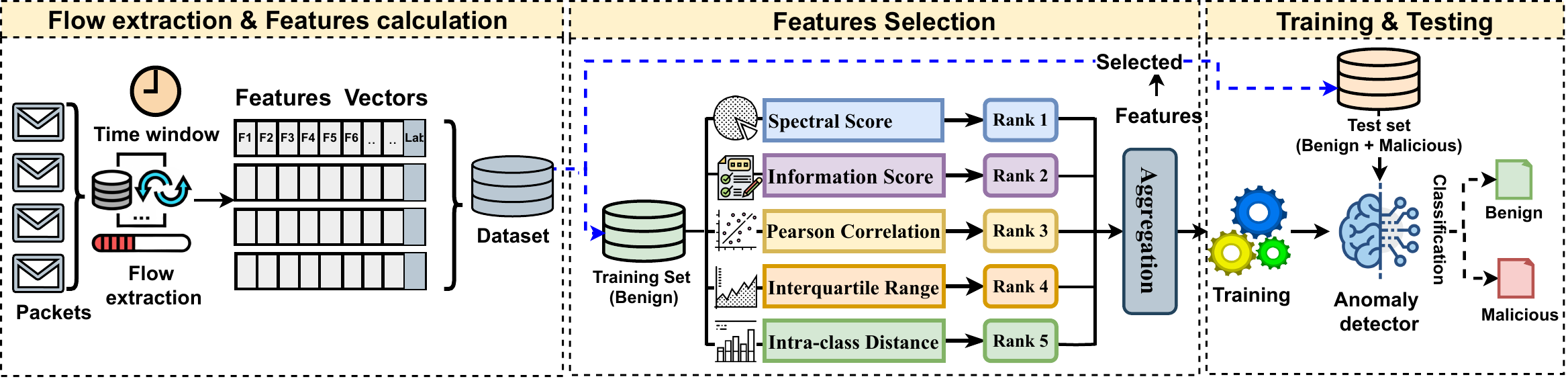}
    \caption{Network Traffic Modeling Workflow}
    \label{fig:workflow}
\end{figure*}

\section{Proposed methodology}
This section outlines the methodology for network traffic analysis, considering both unidirectional and bidirectional flows, as well as packet formats. It explores the features critical for representing traffic and discusses the selection process for identifying those that best characterize the traffic patterns. Finally, it delves into the modeling of traffic using different semi-supervised learning techniques. Figure \ref{fig:workflow} illustrates the workflow of our network traffic analysis methodology.
\subsection{Network traffic representation} \label{sec:fe}
\subsubsection{Flow and features extraction}
We consider both packet-based and flow-based network traffic formats for botnet detection. To identify traffic flows, we utilize 5-tuple information, which includes the source port number, destination port number, and protocol, but exclude source and destination IP addresses to protect user privacy. For each extracted flow, we calculate a set of features within a specified time window, applying the same feature set for packet-based analysis as well. Detailed descriptions of the features calculated for bidirectional flows are available in \cite{CICFlowMeter}, while in-depth explanations for unidirectional flows and packet-based analysis can be found in \cite{tranalyzer}. To further ensure user privacy and avoid biased outcomes, we also exclude payload details, source and destination ports, and timestamps. The network features fall into the following  four main categories:
\begin{itemize}
        \item \textbf{Packet-Based}: This category focuses on metrics related to individual packets, such as their counts and transmission rates. It is crucial for evaluating the volume and flow patterns within the network, providing insights into traffic density and potential congestion points.
    \item \textbf{Byte-Based}: These features examine the volume of data transmitted, encompassing total bytes and sizes of packets. They are key to assessing the network's load and utilization, helping to identify significant data transfers and understand bandwidth consumption.
    \item \textbf{Time-Based}: Time-based metrics capture the temporal characteristics of network traffic, including flow durations and the intervals between packet transmissions. They offer valuable perspectives on the timing of network activities, highlighting patterns of usage and detecting irregular or anomalous behaviors.
    \item \textbf{Protocol-Based}: Derived from specific protocol information, such as TCP/UDP protocols and various header details, this category is instrumental in distinguishing types of network traffic. 
\end{itemize}

\subsubsection{Features selection}
Our objective is to refine the features set, retaining only those features that are truly pertinent. Considering that this study adopts a realistic configuration, where we predominantly have access to normal traffic and very little to no malicious traffic, we require a feature selection technique that is capable of selecting attributes with a single class (normal). In this study, we implement the filter feature selection methodology outlined by Lorena et al.~\cite{lorena}, which aligns exceptionally well with our specific requirements. To select features, five essential criteria for individually evaluating and prioritizing features are employed, detailed as follows:
\begin{itemize}
    \item \textbf{Spectral Score (SPEC)}: It involves constructing a similarity matrix \(S\) for all data pairings, using the Radial Basis Function (RBF) to compute similarities between two instances \(x_{i}\) and \(x_{j}\), as formulated as: \begin{equation}
        \label{eq1}
        S_{ij}=\mathrm{e}^{-\frac{{\parallel x_{i}-x_{j} \parallel}^2}{2 \sigma^{2}}}
    \end{equation}

    \item \textbf{Information Score}: this score aims to maximize information gain for the target class. The randomness within the data is gauged using entropy calculations based on the RBF similarity matrix \(S\), as depicted in Equation~\ref{eq2}:
    \begin{equation}
        \label{eq2}
        E=-\sum_{i=1}^n \sum_{j=1}^nS_{ij}\log_{2}S_{ij}+(1-S_{ij})\log_{2}(1-S_{ij})
    \end{equation}

    \item \textbf{Pearson Correlation}: It calculates the Pearson correlation between each feature and the others, summing the absolute values of these correlations, as shown in Equation~\ref{eq3}:
    \begin{equation}
        \label{eq3}
        corr(f_{i})=\sum_{j=2}^m|pearson(f_{i},f_{j})|
    \end{equation}

    \item \textbf{Intra-class Distance}: This metric quantifies the average distance of all instances within a class from the class centroid (\(\bar{x}\)), as illustrated in Equation~\ref{eq4}:
    \begin{equation}
        \label{eq4}
        IE=\frac{1}{n}\sum_{i=1}^nd(x_{i},\bar{x})
    \end{equation}

    \item \textbf{Interquartile Range}: This is calculated from the feature distribution within the target class, focusing on the interquartiles. The Interquartile Range (IQR) is calculated as follows:
\begin{equation}
    \text{IQR} = Q3 - Q1
\end{equation}
where \(Q1\) and \(Q3\) respectively represent the first and third quartiles of the data set.
\end{itemize}
To synthesize their outcomes of these metrics , three ranking aggregation methods are proposed: mean, majority, and Borda, with the mean method—averaging the feature positions across ranking lists—being selected for our study. 
\subsection{Botnet Network traffic detection}\label{SOL}
Our methodology is designed to accurately model the baseline network traffic of connected devices, identifying any deviations from established patterns as potential security threats. To accurately model the normal network traffic pattern of IoT devices, we rigorously assess the feasibility and effectiveness of our approach by scrutinizing the following five prominent semi-supervised learning techniques. The selected algorithms exemplify varied semi-supervised detection methods: Elliptic Envelope defines geometric boundaries to identify outliers, Isolation Forest employs ensembles for complex pattern detection, Local Outlier Factor uses clustering for nuanced analysis, and Autoencoder leverages neural networks for subtle behavior detection. Brief descriptions of their functions follow.
\begin{itemize}
    \item \textbf{Isolation Forest (IF) \cite{if}}: Utilizes Isolation Trees to detect anomalies, with the anomaly score \(s(x, n)\) reflecting the ease of isolating a point. The score is calculated as:
    \begin{equation}
    s(x, n) = 2^{-\frac{E(h(x))}{c(n)}}
    \end{equation}
    where \(E(h(x))\) is the average path length to isolate the point in iTrees, and \(c(n)\) is a normalization factor based on the dataset size \(n\).

    \item \textbf{Elliptic Envelope (EE) \cite{ee}}: Encloses data points within an ellipse, assuming Gaussian distribution. Outliers are identified using the Mahalanobis distance \(D_M(x)\):
    \begin{equation}
    D_M(x) = \sqrt{(x - \mu)^T S^{-1} (x - \mu)}
    \end{equation}
    Points with \(D_M(x)\) exceeding a threshold are flagged as outliers.

    \item \textbf{Local Outlier Factor (LOF) \cite{lof}}: Identifies outliers by comparing local density. The LOF score for a point \(x\) is given by:
    \begin{equation}
    LOF_k(x) = \frac{\sum_{y \in N_k(x)} \frac{lrd_k(y)}{lrd_k(x)}}{|N_k(x)|}
    \end{equation}
    This score indicates the extent to which a point is an outlier based on its neighborhood density.

    \item \textbf{One-Class SVM (OSVM) \cite{osvm}}: Separates data points from the origin in a high-dimensional space. The optimization problem is:
    \begin{equation}
    \operatorname*{minimize}_{w,\ \xi_i,\ \rho } \frac{1}{2} \lVert w \rVert ^2 + \frac{1}{\nu n} \sum_{i=1}^n \xi_i - \rho
    \end{equation}
    The decision function \(f(x)\) classifies points as normal or anomalous:
    \begin{equation}
    f(x) = \operatorname{sgn}((w \cdot \phi(x)) - \rho)
    \end{equation}

    \item \textbf{Deep Autoencoders (AE)}: Detects anomalies through the reconstruction error \(E(x)\):
    \begin{equation}
    E(x) = ||x - \hat{x}||^2
    \end{equation}
    A high reconstruction error indicates an anomaly, due to significant deviation from the normal data pattern.
\end{itemize}


\begin{table}[ht]
\centering
\caption{Sample Distribution Across the Different datasets}
\label{tab:sample_dist}
\begin{tabular}{@{}lcccccc@{}}
\toprule
Dataset            & Normal & DDoS  & Scan  & Attack & C\&C    & Download \\ \midrule
Unidirectional  & 100966  & 95337 & 31818  & 27754  & 8412 & 13       \\
Bidirectional   & 10304  & 11500 & 18370 & 690    & 3650  & 11       \\
Packet          & 10000  & 4500  & 2000  & 2200   & 3368  & 1554     \\ \bottomrule
\end{tabular}
\end{table}

\begin{table*}[h]
\centering
\caption{Predictive performance of different classifiers considering different traffic formats}
\label{tab:comp}
\begin{tabular}{@{}llS[table-format=2.2]S[table-format=2.2]S[table-format=2.2]S[table-format=2.2]S[table-format=2.2]S[table-format=2.2]@{}}
\toprule
\textbf{{Classifier}} & \textbf{{Traffic}} & \textbf{{Precision(\%)}} & \textbf{{Accuracy(\%)}} & \textbf{{Recall(\%)}} & \textbf{{FPR(\%)}} & \textbf{{F1-Score(\%)}} & \textbf{{AUC(\%)}} \\ \midrule
\multirow{3}{*}{One-Class SVM (OSVM)} & Bi. Flow & \textbf{99.98} & \textbf{99.97} & \textbf{99.99} & 15.53 & \textbf{99.98} & 92.23 \\
 & Uni. Flow & 99.79 & 97.08 & 97.18 & 5.75 & 98.47 & \textbf{95.71} \\
 & Packet & 99.92 & 99.92 & \textbf{100.00} & \textbf{1.53} & 99.96 & \textbf{99.23} \\
\multirow{3}{*}{Local Outlier Factor (Lof)} & Bi. Flow & 99.97 & 95.85 & 95.87 & 20.58 & 97.79 & 87.65 \\
 & Uni. Flow & 94.00 & 83.87 & 94.00 & 6.00 & 86.40 & 78.93 \\
 & Packet & 98.74 & 98.78 & \textbf{100.00} & 24.83 & 99.36 & 87.58 \\
\multirow{3}{*}{Isolation Forest (IF)} & Bi. Flow & 99.92 & 76.56 & 76.57 & 34.76 & 82.03 & 70.91 \\
 & Uni. Flow & 95.13 & 81.20 & 95.13 & 4.87 & 85.33 & 77.47 \\
 & Packet & 98.83 & 87.03 & 87.39 & 19.96 & 92.26 & 83.71 \\
\multirow{3}{*}{Elliptic Envelope (EE)} & Bi. Flow & 99.97 & \textbf{99.82} & \textbf{99.85} & 20.97 & \textbf{99.91} & 89.44 \\
 & Uni. Flow & 88.00 & 63.00 & 88.00 & 12.00 & 72.20 & 59.87 \\
 & Packet & 98.94 & 98.98 & \textbf{100.00} & 20.78 & 99.47 & 89.61 \\
 \multirow{3}{*}{Autoencoder (AE)} & Bi. Flow & 98.45 & 42.11 & 36.88 & 5.79 & 53.66 & 65.55 \\
 & Uni. Flow & 99.74 & 97.05 & 97.23 & 10.10 & 98.47 & 93.56 \\
 & Packet & 97.87 & 97.13 & 98.51 & 8.07 & 98.19 & 95.22 \\\bottomrule
\end{tabular}
\end{table*}

\section{Performance Evaluation}\label{SIM}
\subsection{Datasets generation}
Our study leveraged the Aposemat IoT-23 dataset \cite{iot23}, sourced from the Stratosphere Laboratory at CTU University, Czech Republic. This dataset contains twenty-three scenarios of IoT network traffic, including real malware infections and benign traffic.  Due to the IoT-23 dataset's vast size, we couldn't analyze it in full. Instead, we chose representative samples from each scenario to capture the diversity of bot malware and activities. We developed scripts leveraging two distinct traffic exporters to process network flows: CICFlowMeter \cite{CICFlowMeter} for extracting bidirectional flows and Tranalyzer \cite{tranalyzer} for unidirectional flow extraction. For comprehensive details on the features computed for bidirectional flows, the reader is directed to \cite{CICFlowMeter}. Similarly, in-depth analyses concerning unidirectional flows and packet-based assessments are thoroughly documented in \cite{tranalyzer}. From the initial dataset consisting of PCAP captures, we meticulously curated three distinct datasets: one containing bidirectional flows, a second comprising unidirectional flows, and a third dedicated to packet-level data. The distribution of samples by dataset and type, following data preprocessing operations such as cleaning, converting categorical attributes to numerical format, and normalization, is depicted in Table \ref{tab:sample_dist}.

\subsection{Experimental results}
We employed the Scikit-learn package for the implementation of the anomaly detection models. These models were trained and tested within the Google Colab cloud environment. We utilized a random search, a lightweight and effective method, to identify the optimal combination of hyperparameters. Additionally, we conducted a 5-fold cross-validation to ensure the robustness of our models. The model evaluation was based on the performance metrics detailed below: 
\begin{itemize}
    \item Precision: $\frac{TP}{TP + FP}$
    \item Accuracy: $\frac{TP + TN}{TP + FN + FP + TN}$
    \item Recall: $\frac{TP}{TP + FN}$
    \item FPR (False Positive Rate): $\frac{FP}{FP + TN}$
    \item F1-Score: $F1\_Score = \frac{2 \times TP}{2 \times TP + FP + FN}$
    \item AUC (Area Under the ROC Curve): Measures the entire two-dimensional area underneath the entire ROC curve from (0,0) to (1,1)
\end{itemize}
TP, TN, FP, and FN denote true positive, true negative, false
positive, and false negative, respectively.


\begin{table*}[htbp]
\centering
\caption{Evaluation of the predictive performances by varying the time-window size}
\label{tab:tw}
\begin{tabular}{@{}lcccccc@{}}
\toprule
\textbf{TW (s)}    & \textbf{Precision (\%)} & \textbf{Accuracy (\%) }& \textbf{Recall (\%)} & \textbf{FPR (\%) }  & \textbf{F1-Score (\%)} & \textbf{AUC (\%)}   \\ \midrule
Default   & 99.79   & 97.08  & 97.18 & 5.75 & 98.47 & 95.71 \\
300       & 99.56   & 96.83  & 97.06 & 6.84 & 98.29 & 95.11 \\
100       & 99.36   & 99.36  & 99.93 & 5.61 & 99.65 & 97.16 \\
10        & 98.04   & 98.72  & 99.96 & 3.37 & 98.99 & 98.29 \\
1        & 99.41   & 99.27  & 99.65 & 2.07 & 99.53 & 98.79 \\ \bottomrule
\end{tabular}
\end{table*}
\begin{table*}[htbp]
\centering
\caption{Features selection evaluation}
\label{tab:fs}
\begin{tabular}{@{}lccccccc@{}}
\toprule
\textbf{Features set}     & \textbf{Nb. Features} & \textbf{Precision (\%)} & \textbf{Accuracy (\%)} & \textbf{Recall (\%)} & \textbf{FPR (\%)}  & \textbf{F1-Score (\%)} & \textbf{AUC (\%)}   \\ \midrule
All Features      & 79                & 99.41   & 99.27  & 99.65 & 2.07 & 99.53 & 98.79 \\
Selected Features & 51                & 99.56    & 98.53   & 98.55 & 1.53 & 99.05  & 98.51 \\ \bottomrule
\end{tabular}
\end{table*}

\begin{figure*}
\centering
\begin{subfigure}[b]{0.2\textwidth}
\includegraphics[width=\textwidth]{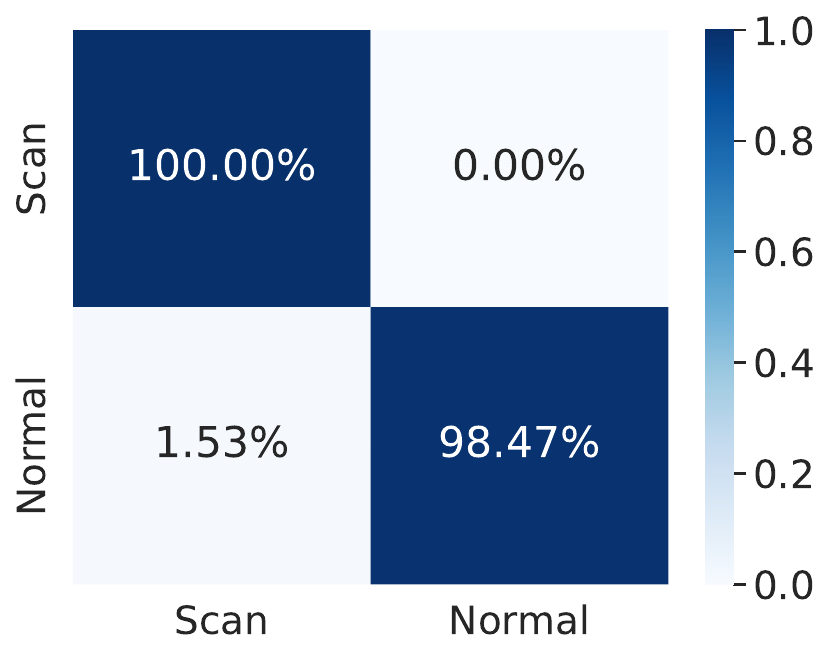}
\caption{Scan}
\end{subfigure}
\begin{subfigure}[b]{0.2\textwidth}
\includegraphics[width=\textwidth]{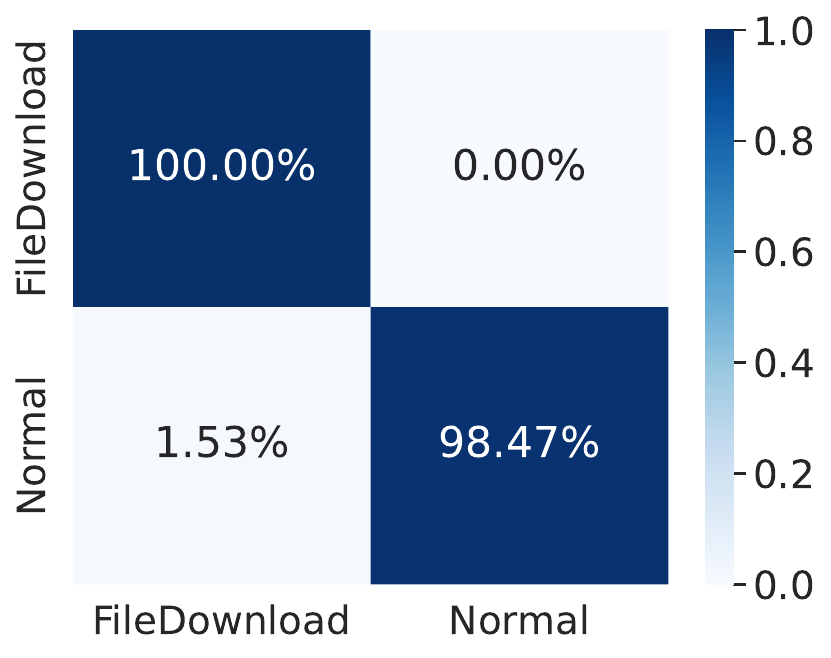}
\caption{File download}
\end{subfigure}
\begin{subfigure}[b]{0.2\textwidth}
\includegraphics[width=\textwidth]{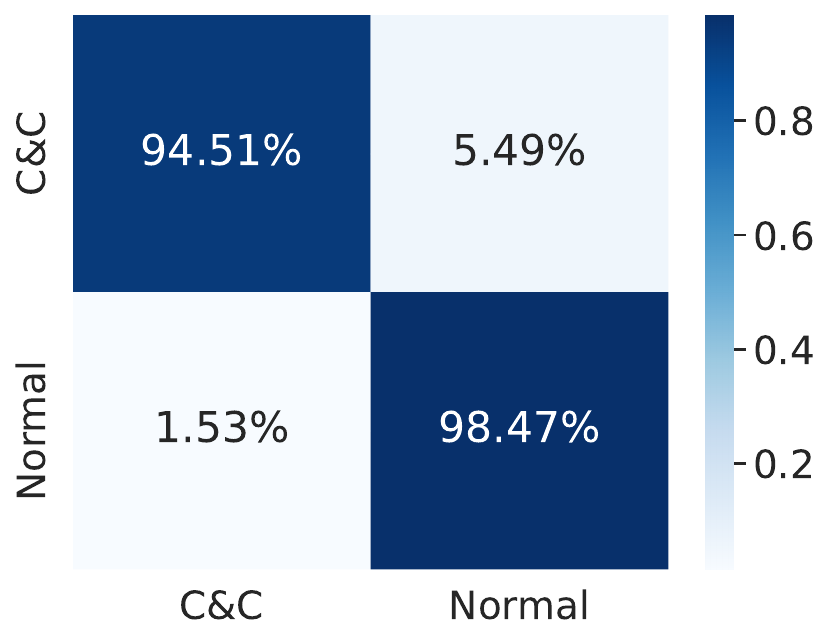}
\caption{C\&C}
\end{subfigure}
\begin{subfigure}[b]{0.2\textwidth}
\includegraphics[width=\textwidth]{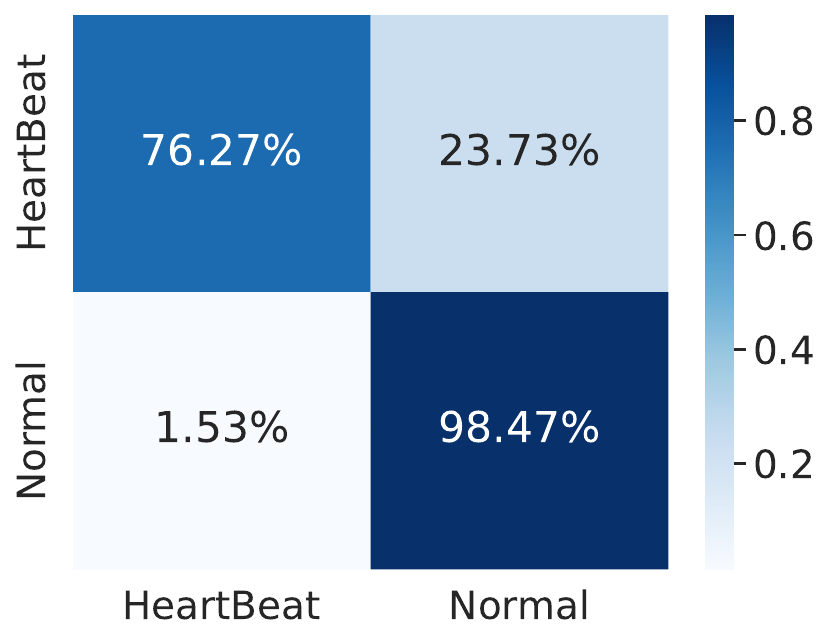}
\caption{Heart Beat}
\end{subfigure}
\caption{Unidirectional Flow-Based Detection Across Various Types of Botnet Traffic}
\label{fig:mat}
\end{figure*}

\subsubsection{Detection performances}
Upon analyzing the performance metrics presented in Table \ref{tab:comp}, OSVM stands out as the best-performing classifier across a variety of traffic types, including bidirectional, unidirectional flows, and packet-based formats. Notably, its performance in packet-based traffic is exceptional, demonstrating a perfect detection rate (recall) alongside a remarkably low false alarm rate (FPR). This efficacy is further evidenced by a high AUC score, underscoring OSVM's capability to accurately differentiate between normal and bot traffic with minimal error. Therefore, the optimal configuration for detecting botnet traffic involves utilizing OSVM with network traffic represented in packet format. For unidirectional flow and packet traffic, the autoencoder is the second-best choice, with its high recall and precision. AE is especially useful in cases where One-Class SVM (OSVM) cannot be used, like in federated learning.


\subsubsection{Detection Delay}
Considering packet-level monitoring, the detection delay is measured in milliseconds, enabling near real-time detection for immediate identification and mitigation of botnet traffic. For unidirectional flow, which demonstrates the second-best performance, we tested various sampling time windows to determine the minimal detection delay that does not compromise the system's performance. Experimenting with different Time Window (TW) durations aimed to find an optimal balance for quick and accurate detection. Contrary to our initial assumption that larger TWs would enhance detection capabilities, as indicated in Table \ref{tab:tw}, a 1-second TW size surprisingly yielded the highest precision, accuracy, and F1-score while maintaining the lowest false positive rate. However, it is important to note that detection delay is influenced by various factors, including the traffic volume, system resources, feature extraction time, and implementation efficiency. In the best-case scenario, the detection delay could be at the lower end of its range, leaning towards milliseconds for packet-based detection, and around 1 second for unidirectional flow-based detection.

\subsubsection{Features selection}
In the development of our anomaly detection framework, the selection of an appropriate feature set is critical to effectively model the normal network pattern. Our analysis has led us to a feature set predominantly composed of Time-Based and Protocol-Based attributes, which together represent an impressive 75.38\% of the total features used---specifically, 36.92\% for Time-Based features and 38.46\% for Protocol-Based features. This composition was chosen based on the premise that the nuances of temporal communication patterns and protocol-specific data are instrumental in establishing a baseline of `normal' traffic. Feature selection, as depicted in Table \ref{tab:fs}, achieved a reduction of 35 \%, bringing the number of features down from 79 to 51. Considering the AUC, there was a very slight decrease; overall, the performance remained the same, even with a reduction of more than one-third of the features. Through this process, the model was efficiently streamlined, preserving its efficacy with a notably reduced set of features.

\subsubsection{Discussion}
Given that packet-based detection achieves a 100\% detection rate for all traffic types, we now examine the performance of detection in the case of unidirectional flow-based detection. Based on the confusion matrices presented in Figure \ref{fig:mat}, we observe perfect detection of scan flows and download inductions. It is particularly noteworthy how we have managed to anticipate the detection of C2 traffic, which poses a greater challenge compared to previous traffic types, achieving an impressive success rate of 94\%. As for Heartbeat traffic, the detection rate is at 76\%, making it the most difficult type of traffic to detect. In the Mirai botnet, HeartBeat is often associated with a type of communication used to maintain connection and check the presence of a bot within the botnet. The heartbeat typically involves periodic, simple messages sent between the bot and the C2 server. For Mirai, these heartbeat messages can be very basic, just a few bytes, to confirm the bot is still active and connected.

\section{Conclusion} \label{CON}
This study has conclusively demonstrated the feasibility of effectively modeling normal network traffic for IoT devices. By leveraging packet-based and unidirectional flow formats, alongside Time-Based and Protocol-Based features, we have optimized the representation of network traffic. The use of semi-supervised learning approaches, especially the One-Class SVM and Autoencoder methods, has been validated for modeling normal IoT traffic patterns. Our results confirm the efficacy of semi-supervised ML techniques in accurately detecting botnet activities, including stealth netowork traffic like scanning and command-and-control (C2) communications. Significantly, the study has not only proven the ability to detect bots at early stages but also achieved a detection delay of less than 1 second in packet-based traffic, with a perfect detection rate and a FPR under 2\%, and a 1-second detection delay in unidirectional flow traffic, attaining a 98\% detection rate with around 2\% FPR. Additionally, we have concluded from this study that, although flow-based detection, widely used by the community, yields good results, it demonstrates inferior performance compared to the packet-based approach when it comes to detecting C2 traffic.






\section*{Acknowledgment}

This work was supported by the 5G-INSIGHT bilateral project (ID: 14891397) / (ANR-20-CE25-0015-16), funded by the Luxembourg National Research Fund (FNR), and by the French National Research Agency (ANR).


\bibliographystyle{unsrt}
\bibliography{ref}

\end{document}